\documentstyle[prb,aps,twoside,multicol,rotate,epsf]{revtex}

\newcommand{\be}{\begin{equation}}
\newcommand{\ee}{\end{equation}}
\newcommand{\ba}{\begin{eqnarray}}
\newcommand{\ea}{\end{eqnarray}}
\newcommand{\ra}{\rangle}
\newcommand{\la}{\langle}

\begin{document}
\title{Calculating the Thermal Rate Constant with Exponential Speed-Up
on a Quantum Computer}
\author{Daniel A. Lidar and Haobin Wang}
\address{Department of Chemistry, The University of California, Berkeley,\\
CA 94720}
\maketitle

\begin{abstract}
It is shown how to formulate the ubiquitous quantum chemistry problem of
calculating the thermal rate constant on a quantum computer. The
resulting exact algorithm scales exponentially faster with the
dimensionality of the system than all known ``classical'' algorithms for
this problem.
\newline
\end{abstract}

\begin{multicols}{2}
\markboth{{Lidar and Wang}}{{Submitted to {\it J. Chem. Phys.},
May 1998}}

\newpage

\section{Introduction}

\label{intro}

It is well known that an exact calculation of the thermal rate constant is a
problem that scales exponentially with the number of degrees of
freedom.\cite{Miller:98} But what if we had a quantum
computer\cite{Feynman:QCBenioff:82Deutsch:85,Deutsch:89} 
(QC) at our disposal?
As we will show here, then the calculation can be speeded up exponentially.
Exponential speed ups on QCs have been demonstrated in a number of problems,
the most famous of which are Shor's algorithm for the factoring
problem,\cite{Shor:94,Shor:97,Ekert:96} and Grover's algorithm for database
search.\cite{Grover:96Grover:97Grover:97aCollins:97Boyer96me:NASA98} In
the context 
of physics problems, exponential speed up 
has been demonstrated mostly in the context of simulation of the many-body
Schr\"{o}dinger
equation.\cite{Zalka:98,Lloyd:96Boghosian:97aBoghosian:97cAbrams:97}
Other physics applications have also been proposed, such as studying quantum
chaos\cite{Schack:98} and Ising spin glasses.\cite{me:PRE97a}
At the 
experimental level QCs are still in a stage of infancy, although very
impressive first steps towards implementation have been taken using
ions in ion traps,\cite{Cirac:95Monroe:95} atoms in high-finesse microwave
cavities,\cite{Turchette:95} and molecular spins in
NMR.\cite{Gershenfeld:97Cory:97aChuang:98b} In particular, Chuang
{\it et al.}, using a chloroform NMR-QC, recently implemented for the first time a quantum algorithm
(Grover's) which outperforms any classical algorithm designed to solve
the same task.\cite{Chuang:98} Another algorithm for which QCs offer
an exponential
speed up compared to classical computers, known as ``Deutsch's
problem,''\cite{Deutsch:92} has also been implemented on a chloroform
NMR-QC,\cite{Chuang:98a} and by Jones {\it et al.} on a cytosine
NMR-QC.\cite{Jones:98a,Jones:98b} These impressive achievements signal
quite clearly, albeit for very simple applications at this point, that
QCs may perhaps sooner than expected play an important role in simulations. For a 
comprehensive introduction to quantum computation, we refer the
interested reader to a number of recent
reviews.\cite{Bennett:95Lloyd:95:1DiVincenzoSteane:98}

The
extensive body of work in quantum computation has, however, to date
not addressed a computational problem of direct relevance to the
quantum chemistry 
community (apart, of course, from the general simulation of the Schr\"{o}dinger
equation). In this work we show how to formulate the ubiquitous quantum chemistry problem of
calculating the thermal rate constant on a QC, and by doing so, how the calculation can be speeded up
exponentially with the number of degrees of freedom. The rate constant
is the single most important number characterizing chemical reactions,
and thus great efforts have been invested in designing efficient
and {\em exact} ``classical'' computational
ways to obtain it. 
Important progress along this line has been made by Miller
{\it et al.,}\cite{wardmiller96} Light {\it et al.,}\cite{light}, and
Manthe {\it et al.,}\cite{manthe} based on the efficient
evaluation of the flux correlation function.\cite{mst,yamamoto}
Approximate methods have also been developed for obtaining the rate
constants. For example, the popular mixed quantum-classical
model\cite{billinggerber,mixed} whereby one integrates the 
time-dependent Schr\"{o}dinger equation for (a few) degrees of
freedom that are treated quantum
mechanically, simultaneously with the classical equations of motion
for the (many) degrees of freedom  that are treated by classical
mechanics; the semiclassical initial value
representation\cite{miller70} (SC-IVR) that has had a rebirth of 
interest\cite{klukivrhellerivrkayivr,brummerivrtannormonoivr,millerivr}
as a way for including quantum effects in molecular 
dynamics simulations; and the (further) linearizing approximation to
the SC-IVR which leads to a much simpler form for the rate
expression.\cite{wang98}  While these methods enjoy favorable
computational scaling properties, they are inherently
approximate and thus not in the context of the present paper.

In spite of these
significant advances, exact classical algorithms can at most achieve a
polynomial speed up in a problem that inherently scales exponentially. Indeed, when classical algorithms are described as $O(N^3)$
instead of $O(N^2)$, it should be remembered that $N$ is itself
exponentially large! The anticipated advance in quantum computation
would therefore have revolutionary consequences for quantum chemistry,
rendering ``classical'' simulation methods essentially obsolete. Of
course, QCs are still at best many years away from reaching the point
of replacing classical computers. Nevertheless, it is of considerable
interest to exhibit an explicit QC algorithm for a problem as central as the
computation of the rate constant, and this is the task we undertake
here. The paper is organized as follows: 
In Sec.~\ref{qc-intro} we introduce some pertinent concepts of
quantum computation. Then, in Sec.~\ref{rc} we briefly rederive the
exact quantum expression for the rate constant. The next sections are
the heart of the paper, where the QC algorithm for calculation of the
rate constant is described in detail. Sec.~\ref{conclusions} concludes.

\section{A Brief Introduction to Quantum Computation}
\label{qc-intro}

Let us denote all the degrees of freedom of the problem at hand by the collective
variable $\vec{q}\equiv (q_{1},...,q_{M})$, and let us assume for simplicity
that after discretization the number of points per degree of freedom is $
2^{l}$. Then the Hilbert space for a given Hamiltonian ${\bf H}\left( \vec{q}
,\vec{p} \right) $ is of dimension 
\begin{equation}
N=2^{\nu },\,\,\,\nu =lM
\end{equation}
whence the exponential scaling. To set up the problem on a QC
one introduces a ``register'' of $\nu $ ``qubits'' (two-level systems),
which can be in a superposition state $|\phi _{i}\rangle =a_{i}|0_i\rangle
+b_{i}|1_i\rangle $ (with $|a_{i}|^{2}+|b_{i}|^{2}=1)$. Each group of $l$
qubits corresponds to one of the degrees of freedom $q_{j}$. Initially the
quantum register $|\Phi \rangle $ is in a direct product state with all
qubits in the $|0\rangle $ state: 
\begin{equation}
|\Phi \rangle =\bigotimes_{i=1}^{\nu }|0_{i}\rangle \text{ .}  \label{eq:0}
\end{equation}
Allowed operations on the register are all the {\em unitary transformations}
(corresponding to propagation of the register), and all {\em measurements},
i.e., projections onto subspaces of the full register Hilbert space ${\cal H}
$. However, by convention the unitary transformations should be explicitly
given in terms of operations on single and two qubits at the most, since it
is such one and two-qubit ``gates'' that one can expect to construct in
practice. Also, allowing arbitrarily large gates would not constitute a
general-purpose computer (this is similar to the situation with classical
computers, where the number of distinct logical elements is a small and
finite set). Simplifying an early construction by Deutsch,\cite{Deutsch:89} it has been proven\cite
{DiVincenzo:95Barenco95:2Sleator:95Lloyd:95} that the set of {\em all
single-qubit gates} [the group $U(2)$] with in addition the ``{\em 
controlled-not}'' (CNOT) 
\begin{equation}
|\epsilon _{1},\epsilon _{2}\rangle \longmapsto |\epsilon _{1},(\epsilon
_{1}+\epsilon _{2})\text{mod}2\rangle \:\:\:\:\: (\epsilon_i=0,1) \,,
\label{eq:CNOT}
\end{equation}
form a {\em universal} set of gates: all unitary transformations, of
arbitrary size, can be constructed using a polynomially large set of the 1-
and 2-qubit gates. Therefore the restriction to this set of gates is
sufficient to simulate {\em any} computable function. This, of course,
includes (and potentially exceeds) anything that is computable on a
classical computer. 

Now, let us see how to set up a superposition state on the quantum register
corresponding to all possible initial classical positions. In the basis $
|0\rangle ={\ {
{1  \choose 0}
} }$ and $|1\rangle ={\ {
{0  \choose 1}
} }$, one applies the one-qubit unitary ``Hadamard transform'' 
\begin{equation}
{\bf R}=\frac{1}{\sqrt{2}}\left( 
\begin{array}{rr}
1 & 1 \\ 
1 & -1
\end{array}
\right) \,,  \label{eq:R}
\end{equation}
which is a $\pi /2$ rotation on the Bloch sphere of each qubit. Thus: 
\ba
|\Phi \rangle \longmapsto |\Phi'\rangle &=& \bigotimes_{i=1}^{\nu } 
{\bf R}|0_{i}\rangle =\bigotimes_{i=1}^{\nu }\frac{(|0_{i}\rangle
+|1_{i}\rangle )}{\sqrt{2}} \nonumber \\
&=& \frac{1}{\sqrt{N}}\sum_{j=0}^{N-1}|j\rangle \,,
\label{eq:PHI}
\ea
where $j$ is the decimal representation of the register state with the
corresponding binary value, so $|j\rangle $ is a convenient shorthand
notation for a tensor product of $\nu $ single-particle
states. Eq.(\ref{eq:PHI}) represents the desired superposition over
all initial positions. For example,
let $l=2$ (so the number of grid points per degree of freedom is $4$) and $
M=2$ (e.g., $q_{1}=x$ and $q_{2}=y$ in a 2-dimensional problem involving a
linear triatomic vibrating molecule) so $N=16$; how is $\vec{q}=(2\Delta
x,3\Delta y)$ represented in the register? In binary, $\vec{q}
=(\{1,0\}\Delta x,\{1,1\}\Delta y)$, so that $j=11$, i.e., $|j\rangle
=|1\rangle \otimes |0\rangle \otimes |1\rangle \otimes |1\rangle $,
corresponds to $x=2\Delta x$ and $y=3\Delta y$. In this way the quantum
register supports a superposition over all discretized values of the degrees
of freedom $\vec{q}$. By linearity, unitary evolution of the register
amounts to parallel propagation on all of the exponentially many grid
points. We will find it convenient to work from now on with $j$ as the
collective degrees of freedom variable, instead of $\vec{q}$. The
transformation between the two is straightforward. Note further that the
register states $|j\rangle $ are {\em position eigenstates}.

Suppose now that the initial wavefunction $|\Psi (0)\rangle $ in our
scattering problem has amplitude $\alpha _{j}$ to be at position $j$. As
explained below, it is in fact not essential to utilize this initial
condition, and an equal superposition of all possible position
eigenstates will be sufficient in most cases. Nevertheless, as shown
by Zalka\cite{Zalka:98} (and will not be repeated here), it is
possible to initialize the register to 
the state  
\begin{equation}
|\Phi'\rangle \longmapsto |\Phi''\rangle
=\sum_{j=0}^{N-1}\alpha _{j}|j\rangle \,,  \label{eq:PHI'}
\end{equation}
by means of a suitable unitary transformation. This then represents $|\Psi
(0)\rangle $ on the QC. The dynamics in the scattering problem
is determined by the unitary propagator ${\bf U}=e^{-i{\bf H}t/\hbar }$ :\ $
|\Psi (t)\rangle ={\bf U}|\Psi (0)\rangle $. The crucial advantage
offered by a QC is that, as will be shown below, it is
possible to {\em efficiently} implement this propagator on the quantum
register, so that: 
\begin{equation}
|\Phi''\rangle \longmapsto |\Phi'''\rangle ={\bf U}|\Phi''\rangle .
\label{eq:U}
\end{equation}
In this way we have set up a 1-1 correspondence between the QC
($|\Phi \rangle $) and the dynamics of the problem of interest ($|\Psi
\rangle $). All the relevant information can be extracted from this
simulation by observing the states of the qubits.

\section{The Thermal Rate Constant via the Flux Correlation Function
Formalism}
\label{rc}

Let us now turn to the scattering problem and define a flux operator 
\begin{equation}
{\bf F=}\frac{i}{\hbar }\left[ {\bf H,h}\left[ s(\vec{q})\right] \right] \,,
\label{eq:F}
\end{equation}
where ${\bf h}$ is the Heaviside function and the condition $s(\vec{q})=0$
defines the dividing surface.

The thermal rate constant is written as the time integral of the
flux-flux autocorrelation function:\cite{mst} 
\begin{equation}
k(T)=\frac{1}{Q_{r}(T)}\int_{0}^{\infty }dt\,C_{f}(t)\,,  \label{eq:k}
\end{equation}
where $Q_{r}(T)$ is the reactant partition function per unit volume, and: 
\begin{eqnarray}
C_{f}(t) &=&\text{Tr}[\,e^{-\beta {\bf H}/2}\,{\bf F}\,e^{-\beta {\bf H}
/2}\,e^{i{\bf H}t/\hbar }\,{\bf F}\,e^{-i{\bf H}t/\hbar }]  \nonumber \\
&=&\text{Tr}[\,{\bf F}\,e^{i{\bf H}t/\hbar -\beta {\bf H}/2}\,\,{\bf F}
\,e^{-i{\bf H}t/\hbar -\beta {\bf H}/2}]  \nonumber \\
&=&\text{Tr}[\,{\bf F}\,e^{i{\bf H}\tau ^{\ast }/\hbar }\,\,{\bf F}\,e^{-i 
{\bf H}\tau /\hbar }]  \label{eq:c}
\end{eqnarray}
and where we have also defined
for convenience the complex ``time'' 
\begin{equation}
\tau =t-i\hbar \beta /2\,.  \label{eq:tau}
\end{equation}
Evaluating the trace in the energy eigen-basis $\{|n\rangle \}$ (with ${\bf 
H }|n\rangle =E_{n}|n\rangle $) we obtain: 
\begin{eqnarray}
C_{f}(t) &=&\sum_{n}\langle n|{\bf F}\,e^{i{\bf H}\tau ^{\ast }/\hbar }\, 
{\bf F}\,e^{-i{\bf H}\tau /\hbar }|n\rangle  \nonumber \\
&=&\sum_{n,m}\langle n|{\bf F}\,e^{i{\bf H}\tau ^{\ast }/\hbar }\,|m\rangle
\langle m|{\bf F}\,e^{-i{\bf H}\tau /\hbar }|n\rangle  \nonumber \\
&=&\sum_{n,m}\,e^{iE_{m}\tau ^{\ast }/\hbar }e^{-iE_{n}\tau /\hbar
}\,\langle n|{\bf F}|m\rangle \langle m|{\bf F}|n\rangle  \nonumber \\
&=&\sum_{n,m}\,e^{-\beta (E_{m}+E_{n})/2}e^{i(E_{m}-E_{n})t/\hbar
}\,|\langle n|{\bf F}|m\rangle |^{2}\,.  \label{eq:12}
\end{eqnarray}
Using the commutator form for ${\bf F}$, Eq.(\ref{eq:F}), we find: 
\begin{eqnarray}
\langle n|{\bf F}|m\rangle &=&\frac{i}{\hbar }\langle n|\left[ {\bf H\,h} 
\left[ s(\vec{q})\right] -{\bf h}\left[ s(\vec{q})\right] {\bf H}\right]
|m\rangle  \nonumber \\
&=&\,\frac{i}{\hbar }(E_{n}-E_{m})\,\langle n|{\bf \,h}\left[ s(\vec{q}) 
\right] |m\rangle \,.  \label{eq:nFm}
\end{eqnarray}
Recall from Eq.(\ref{eq:PHI}) that the quantum register naturally supports a
superposition over position eigenstates. Accordingly, let us represent $
|n\rangle $ in the discretized position basis $\{|j\rangle \}$, which we
from now on identify with the QC's ``computational basis'' $
\{|j\rangle \}$ (indeed, the correspondence is 1-1). Thus, let us expand the
energy eigenstates as: 
\begin{equation}
|n\rangle =\sum_{j=0}^{N-1}a_{j}(n)|j\rangle .
\label{eq:n}
\end{equation}
Clearly, the Heaviside function ${\bf h}\left[ s(\vec{q})\right] $ is
diagonal in this basis, so that from Eq.(\ref{eq:nFm}):

\begin{equation}
\langle n|{\bf F}|m\rangle =\frac{i}{\hbar }(E_{n}-E_{m})
\sum_{j=0}^{N-1}a_{j}^{\ast }(n)a_{j}(m)h\left[ s(j)\right] \,\,.
\end{equation}
Hence, finally: 
\ba
C_{f}(t) &=& \frac{1}{\hbar ^{2}}\sum_{n\neq m}\,e^{-\beta
(E_{m}+E_{n})/2}e^{i(E_{m}-E_{n})t/\hbar }\,(E_{n}-E_{m})^{2}
\nonumber \\
&\times& \left|
\sum_{j=0}^{N-1}a_{j}^{\ast }(n)a_{j}(m)h\left[ s(j)\right] \right| ^{2}\,.
\label{eq:cft}
\ea
Our task is therefore to find an algorithm that calculates the spectrum $
\{E_{n}\}$, and the position amplitudes $\{a_{j}(n)\}_{j=0}^{N-1}$ for each
of the eigenstates $|j\rangle $. With these in hand the rest of the
calculation [summations in Eq.(\ref{eq:cft})] can be efficiently
implemented on a classical computer.

At first sight it appears that there
are two problems:\ (1) The spectrum may contain exponentially many energies;
(2) The number of measurements needed must be exponentially large (even if
the number of energy eigenstates is polynomial)\ since there are
exponentially many position eigenstates $|j\rangle $! Regarding (1), the
summation should indeed extend over a polynomially large number of energies
only. This, however, is reasonable if the spectrum is sufficiently
degenerate and, more importantly, since the exponential decrease due to the
Boltzman factors will effectively eliminate the higher end of the spectrum $
E\gtrsim k_{B}T+$(barrier height) even if it is not degenerate. As for (2),
the question is how to get a {\em reasonable} sample of the distribution.
For example, if the distribution is badly behaved in position space, one
could Fourier transform to momentum space and sample there. At any rate,
this problem is identical to the one faced by any classical simulation,
where one settles for a Monte Carlo sampling of the wavefunction. This is 
{\em not} the source of the classical bottleneck, and we will adopt the same
Monte Carlo approach in the quantum simulation. Note, however, that unlike
the classical case where attention has to be given to the issue of
generating statistically {\em independent} Monte Carlo samples, the quantum
simulation {\em automatically} generates truly independent samples by the
projection postulate.\cite{vonNeumann:55}

How can we efficiently calculate the eigenstates and the spectrum? Solving
the time-{\em in}dependent Schr\"{o}dinger equation (SE) 
\begin{equation}
{\bf H}|\psi \rangle =E|\psi \rangle
\end{equation}
can be done on a QC by transforming the problem to the
time-dependent SE and propagating the dynamics with the unitary
time-evolution operator ${\bf U}=e^{-i{\bf H}t}$. Each energy eigenvalue and
eigenstate can then be obtained via known quantum algorithms, to be detailed
below, in polynomial time.

\section{General Outline of the Algorithm}

We now give the algorithm in general terms, to be defined more precisely in
the next section.

\begin{enumerate}
\item  Prepare a register as in Eq.(\ref{eq:0}), and attach some ``ancilla''
qubits to it, also in the $|{\bf 0}\rangle $ state. These will serve as a
quantum scratch-pad to record the results of intermediate measurements. From
now on we will distinguish between the ``{\em main}'' and ancillary
registers.

\item  If a good guess for the initial wavefunction is known, initialize the
register to it as in Eq.(\ref{eq:PHI'}). Else initialize the register to an
equal superposition. Since the computational basis states are {\em position 
} eigenstates in all likelihood they are not energy eigenstates, so will not
be stationary under the SE dynamics. Thus except if the equal superposition
corresponds to some undesirable position -- such as very high above the
barrier so that dissociation sets in immediately -- it is as good a guess as
any. In fact, any random (but reproducible) initial distribution will do.

\item  ``Propagate'' the register in parallel for a time $t$. This
corresponds to a parallel evolution of all the position eigenstates. The
propagation is done very much in analogy to the classical FFT method,\cite
{Kosloff:83-1Kosloff:88} in particular the split time propagation scheme.
\cite{Feit:82} Namely, the potential part is diagonal and can be implemented
directly, whereas for the kinetic part it is necessary to Fourier transform
to and back from momentum space.

\item  Perform a ``von Neumann'' measurement (see Sec.~\ref{vonNeumann}) on
the {\em ancillary} register using the Hamiltonian (energy) as the
observable. This accomplishes a double purpose:

\begin{enumerate}
\item  It allows to obtain an energy $E_{n}$ by measuring the ancillas.

\item  It provides a means to sample the energy-position amplitudes $a_{j}(n)
$.
\end{enumerate}

\item  Repeat steps 1-4 many times until the distribution is converged to
the desired accuracy for all relevant eigenstates. The number of
required repetitions is proportional to this accuracy.

\item  Calculate (classically) the sums in Eq.(\ref{eq:cft}).
\end{enumerate}

\section{The Algorithm in Detail}

\label{algo}

\subsection{Initialization}

Here the register is initialized to the state $|\Phi \rangle
=\bigotimes_{i=1}^{2\nu }|0_{i}\rangle $, where the last $\nu $ qubits are
ancillas. The physics of this initialization step depends on the QC implementation. One conceivable way is cooling to the ground state.

\subsection{Inputting the Initial Wavefunction}

If necessary one inputs the initial wavefunction by the technique of
Zalka.\cite{Zalka:98} Else one employs the Hadamard rotations
technique to create an equal 
superposition over position states, as in Eq.(\ref{eq:PHI}). 
In the former case the register will be in the state: 
\begin{equation}
|\Phi''\rangle =\left( \sum_{j=0}^{N-1}\alpha _{j}|j\rangle
\right) \bigotimes_{i=\nu }^{2\nu }|0_{i}\rangle .
\label{eq:anc}
\end{equation}
In the latter case all $\alpha _{j}=1$.

\subsection{Quantum Propagation Algorithm}

This subsection is the heart of the algorithm; it builds on the approach of
Zalka.\cite{Zalka:98} Assume for simplicity that we have a single particle
of mass $m$ in an external potential $V(\vec{q})$. The full Green's
function for arbitrary time $t$ is:

\begin{equation}
G(x_{1},x_{2}; t)= \la x_1 | e^{-i {\bf H} t/\hbar} | x_2 \ra \,.
\end{equation}
For short time steps
$\Delta t\ll 1/E$
($E$ is a typical energy of the system) this becomes approximately:

\begin{equation}
G(x_{1},x_{2};\Delta t)=\kappa \,\exp \left[ im\frac{\left(
x_{1}-x_{2}\right) ^{2} 
}{2\Delta t}-iV(x_{1})\Delta t\right] \,,
\end{equation}
where $\kappa $ is a normalization factor. Applying this to the amplitudes
is equivalent to acting on the basis states with the inverse transformation.
Thus the position eigenstates, properly normalized, transform as: 
\ba
|j\rangle &\longmapsto& {\bf U}|j\rangle = \nonumber \\
&\frac{1}{\sqrt{N}}&
\sum_{j'=0}^{N-1}\exp \left[ -im\frac{\left(
j-j'\right) ^{2}\Delta x^{2}}{ 2\Delta t}+iV(j\Delta x)\Delta t 
\right] |j'\rangle \,. \nonumber \\
\label{eq:trans}
\ea
This is carried out in parallel on the entire superposition $
\sum_{j=0}^{N-1}|j\rangle $. Suppose the time-step and spatial resolution are
adjusted so that: 
\ba
\frac{m\,\Delta x^{2}}{\Delta t}=\frac{2\pi }{N}\,.  \label{eq:N}
\ea
Then by expanding the exponent Eq.(\ref{eq:trans}) can be written as a
succession of a diagonal transformation, Fourier transform, and another
diagonal transformation, all unitary:

\begin{equation}
{\bf U}|j\rangle =\exp \left[ iF_{2}(j)\right] {\cal F}(j,j')\exp 
\left[ iF_{1}(j')\right] |j'\rangle \,,  \label{eq:Uj}
\end{equation}
where:

\begin{eqnarray}
F_{1}(j)&=& -\pi \frac{j^{2}}{N}  \nonumber \\
F_{2}(j)&=& -\pi \frac{j^{2}}{N}+V(j\Delta x)\Delta t  \nonumber \\
{\cal F}(j,j')|j\rangle &=& \frac{1}{\sqrt{N}}\sum_{j'=0}^{N-1}\exp
\left[ 2\pi i\frac{jj'}{N}\right] 
|j'\rangle \,.  \label{eq:fourier}
\end{eqnarray}
Eq.(\ref{eq:N}) tells us how many qubits $\nu =\log _{2}N$ are needed for
given $\Delta x$ and $\Delta t$:

\begin{equation}
\nu =\log _{2}\frac{2\pi \,\Delta t}{m\,\Delta x^{2}}\,.
\end{equation}
The special form of Eq.(\ref{eq:Uj}), involving diagonal transformations and a
Fourier transform, is due to the structure of the Hamiltonian operator as a
sum of operators diagonal in coordinate and momentum space. As mentioned
above, this is very similar to the situation that arises in the classical
FFT method for solving the SE.\cite{Kosloff:83-1Kosloff:88}

\subsubsection{Diagonal Transformations:}

Consider first executing the diagonal unitary transformations $|j\rangle
\longmapsto \exp \left[ iF(j)\right] |j\rangle $, which can be done as
follows, using the ancillary register [Eq.(\ref{eq:anc})], in the state $| 
{\bf 0}\rangle \equiv \bigotimes_{i=\nu }^{2\nu }|0_{i}\rangle $. The number 
$\nu $ of qubits in this register depends on the accuracy with which $F$
needs to be evaluated (see immediately below). Then the following steps are
applied:

\begin{enumerate}
\item  $|j,{\bf 0}\rangle \longmapsto |j,F(j)\rangle $: evaluation of $F$
and storage of the result in the ancillary register;

\item  $|j,F(j)\rangle \longmapsto \exp \left[ iF(j)\right] |j,F(j)\rangle $
: introducing the phase;

\item  $\exp \left[ iF(j)\right] |j,F(j)\rangle \longmapsto \exp \left[ iF(j)
\right] |j,{\bf 0}\rangle $:\ inversion of step 1 in order to clear the
ancillary register.
\end{enumerate}

Step 1 requires that it is possible to evaluate an arbitrary function and
store the result. This is very similar to the equivalent classical problem,
for which algorithms are known using just the elementary classical gates.
The same can be done in the quantum case, by breaking up the evaluation into
elementary arithmetic operations, for which quantum algorithms have been
designed.\cite{Shor:94,Vedral:96Beckman:96} We will not dwell on this issue
here. Step 3 is just the reverse of step 1 and can therefore be implemented
by running the inverse unitary transformation.

Step 2 has no classical analogue since it involves phases. It can be
implemented if one knows how to do $|x\rangle \longmapsto \exp \left[ ix
\right] |x\rangle $. This can be done by simple single-qubit phase-shifts.
Let $\nu =2k$. Using a binary expansion\ $x=\sum_{l=-k}^{k-1}x_{l}2^{l}$, we
have:\ $|x\rangle =|x_{-k}\rangle \otimes |x_{-k+1}\rangle \otimes \cdots
\otimes |x_{k-1}\rangle $, where $x_{l}=0,1$. In the standard basis $
|0\rangle ={
{1 \choose 0}
}$, $|1\rangle ={
{0 \choose 1}
}$, consider the following unitary operation: 
\begin{equation}
{\bf Q=}\bigotimes_{l=-k}^{k-1}\left( 
\begin{array}{cc}
1 & 0 \\ 
0 & e^{i2^{l}}
\end{array}
\right) \,.  \label{eq:QQ}
\end{equation}
The $l^{\text{th}}$ $2\times 2$ matrix is a unitary operation in the Hilbert
space of qubit number $l$. Thus: 
\begin{equation}
\left( 
\begin{array}{cc}
1 & 0 \\ 
0 & e^{i2^{l}}
\end{array}
\right) |x_{l}\rangle =e^{ix_{l}2^{l}}|x_{l}\rangle \,.
\end{equation}
Therefore the full result is: 
\begin{equation}
{\bf Q}|x\rangle =\bigotimes_{l=-k}^{k-1}e^{ix_{l}2^{l}}|x_{l}\rangle
=e^{i\sum_{l=-k}^{k-1}x_{l}2^{l}}\bigotimes_{l=-k}^{k-1}|x_{l}\rangle
=e^{ix}|x\rangle \,,
\end{equation}
as required.

\subsubsection{The Quantum Fourier Transform}

The quantum Fourier transform (QFT)\ algorithm has been discussed
extensively,\cite
{Shor:97,Cleve:97,Coppersmith:94Griffiths:96,Hoyer:97Jozsa:97Ekert:98}
and some beautiful connections to group theory have been made.\cite
{Hoyer:97Jozsa:97Ekert:98} In view of its central importance in the
algorithm for solving the SE (and indeed in {\em all} efficient quantum
algorithm found so far!), we present a brief derivation here, using the
approach of Cleve {\it et al.}\cite{Cleve:97}

The QFT was defined in Eq.(\ref{eq:fourier}). Using the binary-decimal
notation $j/N=0.j_{1}j_{2}...j_{\nu }$ (recall that $N=2^{\nu }$) where $
j_{1}=0,1$ etc., we note first that:

\ba
e^{2\pi i\,\,jj'/2^{\nu
}} &|& j_{1}',j_{2}',...,j_{\nu }'\rangle =
e^{2\pi i\,\,(0.j_{\nu })j_{1}'}|j_{1}'\rangle  \nonumber \\
&\otimes& e^{2\pi i\,\,(0.j_{\nu -1}j_{\nu
})j_{2}'}|j_{2}'\rangle \otimes
\cdots \otimes e^{2\pi
i(0.j_{1}j_{2} ... j_{\nu })j_{\nu }'}|j_{\nu }'\rangle
\,, \nonumber \\
\ea
It follows that:

\begin{eqnarray}
\sum_{j'=0}^{N-1}\exp \left[ 2\pi i\frac{jj'}{N}\right]
|j'\rangle &=&  \nonumber \\
\left( |0\rangle + e^{2\pi i(0.j_{\nu })}|1\rangle \right) &\otimes& \left(
|0\rangle +e^{2\pi i(0.j_{\nu -1}j_{\nu })}|1\rangle \right) \otimes \cdots
\nonumber \\
&\otimes& \left( |0\rangle +e^{2\pi i(0.j_{1}j_{2}...j_{\nu })}|1\rangle
\right) \,,  \label{eq:sum}
\end{eqnarray}
by expanding out the product on the
right-hand-side and a term-by-term comparison. Thus the
Fourier-transformed state in Eq.(\ref{eq:sum}) is in
fact an ``unentangled'' direct product. This fact greatly simplifies
the implementation of the QFT.

To perform the QFT, one first applies a Hadamard rotation [Eq.(\ref{eq:R})] to $|j_{1}\rangle $
(the first qubit of $|j\rangle $), with the result: 
\begin{equation}
{\bf R}|j_{1}\rangle =\left( |0\rangle +(-1)^{j_{1}}|1\rangle \right)
=\left( |0\rangle +e^{2\pi i(0.j_{1})}|1\rangle \right) \,,
\end{equation}
so: $|j\rangle \longmapsto \left( |0\rangle +e^{2\pi i(0.j_{1})}|1\rangle
\right) |j_{2},...,j_{\nu }\rangle $. Let us now define a new single-qubit
operation, similar to ${\bf Q}$ from Eq.(\ref{eq:QQ}): 
\begin{equation}
{\bf Q}_{l}=\left( 
\begin{array}{cc}
1 & 0 \\ 
0 & e^{2\pi i/2^{l}}
\end{array}
\right) \,.  \label{eq:Q}
\end{equation}
This operation is applied on the first qubit $|j_1\ra$, subject to a
{\em control} by a second qubit $|j_l\ra$
(which itself does not change): a ``controlled rotation''. Namely, if $
j_{l}=0$ one does nothing, if it is $1$, one applies ${\bf Q}_{l}$. This can be
written as the following unitary transformation in the 4-dimensional Hilbert
space of the two qubits, in the standard basis $|j_{1}j_{l}\rangle
=|00\rangle =(1,0,0,0)$, $|01\rangle =(0,1,0,0)$, $|10\rangle =(0,0,1,0)$, $
|11\rangle =(0,0,0,1)$: 
\begin{equation}
C{\bf Q}_{l}=\left( 
\begin{array}{ccc}
1 & 0 &  \\ 
0 & 1 &  \\ 
&  & {\bf Q}_{l}
\end{array}
\right) \,.
\end{equation}
After applying $C{\bf Q}_{2}$ one obtains: 
\begin{equation}
\left( |0\rangle +e^{2\pi i(0.j_{1}j_{2})}|1\rangle \right) \,.
\end{equation}
Next a ``controlled-${\bf Q}_{3}$'' is applied, yielding: 
\begin{equation}
\left( |0\rangle +e^{2\pi i(0.j_{1}j_{2}j_{3})}|1\rangle \right) \,.
\end{equation}
Clearly, this process will eventually generate the desired phase in the
superposition state of the first qubit [corresponding to the last qubit in
Eq.(\ref{eq:sum})]: 
\begin{equation}
\left[ \left( \prod_{l=2}^{\nu }C{\bf Q}_{l}\right) {\bf R}\right]
_{1}|j_{1}\rangle =\left( |0\rangle +e^{2\pi i(0.j_{1}j_{2}...j_{\nu
})}|1\rangle \right) \,.
\end{equation}
where the terms in the product from here onwards are applied {\em low index
first}.

Now we turn to the second qubit. Again, a Hadamard rotation on it has the
effect of: ${\bf R}|j_{2}\rangle =\left( |0\rangle +e^{2\pi
i(0.j_{2})}|1\rangle \right) $. This is followed by a controlled-${\bf Q}
_{2}$, conditioned upon $|j_{3}\rangle $: $\left( |0\rangle +e^{2\pi
i(0.j_{2})}|1\rangle \right) \longmapsto \left( |0\rangle +e^{2\pi
i(0.j_{2}j_{3})}|1\rangle \right) $. After the full operation on $
|j_{2}\rangle $ one obtains: 
\begin{equation}
\left[ \left( \prod_{l=2}^{\nu -1}C{\bf Q}_{l}\right) {\bf R}\right]
_{2}|j_{2}\rangle =\left( |0\rangle +e^{2\pi i(0.j_{2}j_{3}...j_{\nu
})}|1\rangle \right) |j_{2}\rangle \,.
\end{equation}
which corresponds to the one before last qubit in Eq.(\ref{eq:sum}).

The method to generate the entire product in Eq.(\ref{eq:sum}) should now be
clear; collecting all the transformations yields: 
\begin{eqnarray}
|j\rangle &\longmapsto& \prod_{p=1}^{\nu -1}\left[ \left( \prod_{l=2}^{\nu
-p}C {\bf Q}_{l}\right) {\bf R}\right] _{p}|j_{1},...,j_{\nu }\rangle
= \nonumber \\
\left( | 0 \rangle \right. \!\!\! &+& \left. \!\!\! e^{2\pi i(0.j_{1}j_{2}...j_{\nu })}|1\rangle \right)
\otimes \cdots \nonumber \\
&\otimes& \left( |0\rangle +e^{2\pi i(0.j_{\nu -1}j_{\nu
})}|1\rangle \right) \otimes \left( |0\rangle +e^{2\pi i(0.j_{\nu
})}|1\rangle \right) \,. \nonumber \\
\end{eqnarray}
Up to an unimportant bit reversal (which can easily be rectified by
permuting the role of the qubits in the transformations above), this is
exactly the desired result. In other words, the QFT is simply: 
\begin{equation}
{\cal F}=\prod_{p=1}^{\nu -1}\left[ \left( \prod_{l=2}^{\nu -p}C{\bf Q}
_{l}\right) {\bf R}\right] _{p}\,.  \label{eq:QFT}
\end{equation}
This will be applied in parallel, by virtue of the superposition principle,
on all position eigenstates $|j\rangle $. Most importantly, the number of
operations (single- and two-qubit) needed to implement the QFT is seen to be
a mere $\nu (\nu -1)/2$. This is to be compared to the $\nu 2^{\nu }$\
operations required classically, and as emphasized above, is the ``secret''
behind the quantum speedup.

\subsection{von Neuman Measurements}
\label{vonNeumann}

Combining Eqs.(\ref{eq:U}) and (\ref{eq:trans}), at this point the register
is in the state 
\begin{eqnarray}
|\Phi'''\rangle  &=&\sum_{j}\alpha _{j}{\bf U}
|j\rangle = \sum_{j'}\psi _{j'}(t)|j'\rangle   \nonumber \\
\psi _{j'}(t)&=&\sum_{j}\alpha _{j}G^{-1}(j,j';t).
\label{eq:sup}
\end{eqnarray}
A parallel propagation has occurred on all the position eigenstates. By
measuring the qubits one by one, i.e., projecting onto a random position
eigenstate $|j'\rangle $, and repeating this process many times
while collecting the statistics, one can sample the electronic density
function $|\psi _{j'}(t)|^{2}$. Our goal was to find the 
{\em energy}-spectrum and energy-position amplitudes $a_{j}(n)$, so
these should be obtained from the simulation. This can be done using
the so-called ``von Neuman measurement'' trick.\cite{Zalka:98} We will
require an additional propagation step.

A ``measurement apparatus'' that can be made to interact with the QC
is introduced, and is assumed to be equivalent to a 1-dimensional 
quantum mechanical particle. That is, its Hilbert space is spanned by the
basis vectors $|x\rangle $, $x$ real, with ${\bf X}|x\rangle =x|x\rangle $.
In practice this will be another ancillary quantum register, consisting of,
say, $K$ qubits. Now, let us expand the position
eigenstates $|j'\rangle $ in terms of the complete set of energy
eigenstates [recall Eq.(\ref{eq:n})]:

\begin{equation}
|j'\rangle =\sum_{n}a^*_{j'}(n)|n\rangle.
\label{eq:j'}
\end{equation}
Consider next the joint evolution of an energy eigenstate $|n\rangle $ and
the apparatus state $|x\rangle $ ($x$ is arbitrary), under the unitary
operator $\tilde{{\bf U }}=\exp (i{\bf H}$ ${\bf P}t/\hbar )$, where $[{\bf 
X,P}]=i\hbar$. Here ${\bf H}$ acts on the main register and ${\bf X}, {\bf P}
$ act on the apparatus, so $[{\bf X,H}]=[{\bf P,H}]=0$. We will shortly
discuss the implementation of $\tilde{{\bf U}}$. Consider first a formal
Taylor expansion of $\exp (i{\bf H}$ ${\bf P}t/\hbar )$, which yields:

\begin{equation}
\tilde{{\bf U}}|n\rangle |x\rangle =\sum_{l=0}^{\infty }\frac{1}{l!}
(tE_{n})^{l}|n\rangle \,\frac{\partial ^{l}}{\partial x^{l}}|x\rangle
=|n\rangle |x+tE_{n}\rangle .
\label{eq:Utilde}
\end{equation}
Thus $\tilde{{\bf U}}$ does not change the energy eigenstate, but has the
effect of ``shifting the dial $x$'' by an amount proportional to the energy $
E_{n}$. The effect on the position eigenstate $|j'\rangle $ will be:

\begin{equation}
\tilde{{\bf U}}|j'\rangle |x\rangle =\sum_{n}a^*_{j'}(n)
|n\rangle |x+tE_{n}\rangle , 
\end{equation}
and the effect on the full superposition of Eq.(\ref{eq:sup}) is:

\begin{eqnarray}
\tilde{{\bf U}}|\Phi'''\rangle |x\rangle
&=&\sum_{j'}\psi _{j'}(t)\sum_{n} a^*_{j'}(n)
|n\rangle |x+tE_{n}\rangle  \nonumber \\ 
&=&\sum_{n}\xi _{n}(t)|n\rangle |x+tE_{n}\rangle  \nonumber \\
\xi_{n}(t) &=& \sum_{j'} a^*_{j'}(n) \psi _{j'}(t).
\label{eq:ap}
\end{eqnarray}
Now suppose we {\em observe} the state of the apparatus. From
Eq.(\ref{eq:ap}) it is clear that the apparatus has become entangled
with the QC, and by performing the observation the superposition will collapse
onto a particular state $|m\rangle |x+tE_{m}\rangle $. This
happens with probability $|\xi _{m}(t)|^{2}$. Recall that $|x+tE_{m}\rangle $
is represented in binary by the qubits of the apparatus. Since $t$ is a
parameter of the simulation and $x$ is known, all that remains is to measure
the apparatus qubit by qubit, to obtain the energy eigenvalue $E_{m}$! The
accuracy with which these numbers are obtained is proportional to the number
of simulation steps.\cite{Zalka:98}

To implement $\tilde{{\bf U}}$ it is necessary to Fourier transform the $
|x\rangle$ register, just like in the classical FFT case. Specifically, let
us define the Fourier transform pair:

\begin{eqnarray}
{\cal F}|x\rangle &=& |p\rangle = \frac{1}{\sqrt{2^K}} \sum_{x=0}^{2^K-1}
e^{-i x p/\hbar} |x\rangle  \nonumber \\
{\cal F}|p\rangle &=& |x\rangle = \frac{1}{\sqrt{2^K}} \sum_{p=0}^{2^K-1}
e^{i x p/\hbar} |p\rangle .
\end{eqnarray}
Then starting from the initial apparatus state $|x\rangle$, $\tilde{{\bf U}}$
can be implemented as follows:

\begin{eqnarray}
|n\rangle |x\rangle  &\stackrel{{\cal F}(x)}{\longmapsto }&|n\rangle
|p\rangle   \nonumber \\
&\stackrel{\tilde{{\bf U}}}{\longmapsto }&e^{iE_{n}pt/\hbar }|n\rangle
|p\rangle   \nonumber \\
&\stackrel{{\cal F}^{-1}(p)}{\longmapsto }&|n\rangle \frac{1}{\sqrt{2^{K}}}
\sum_{p=0}^{2^{K}-1}e^{i(x+E_{n}t)p/\hbar }|p\rangle   \nonumber \\
&=&|n\rangle |x+E_{n}t\rangle ,
\end{eqnarray}
in agreement with Eq.(\ref{eq:Utilde}).

\subsection{Extracting the Amplitudes from the Measurements}

Note further that after observation of the apparatus, the state of the main
register has been projected onto $|m\rangle $, an energy eigenstate. The $
\tilde{{\bf U}}$ propagation had a remarkable outcome: it transformed the
information in the main register from a superposition over position
eigenstates $|j'\rangle $ to one over energy eigenstates $|n\rangle 
$. Unfortunately, apart from utilizing this in a subsequent evolution,
this, however, does 
not appear to be particularly useful, for it is not clear how the energy
eigenstates are enumerated in the main register after the von Neuman
propagation. Nevertheless, it is possible to obtain the amplitudes $a_{j}(n)$
needed to complete the calculation in Eq.(\ref{eq:cft}). Note first
that by Eq.(\ref{eq:ap}):

\begin{equation}
\sum_{n}\xi _{n}(t)a_{j}(n)=\psi _{j}(t).
\label{eq:a}
\end{equation}
Now, the
simulation yields, by performing the whole procedure a sufficient
number of times, an estimate of the {\em probabilities} $|\psi
_{j}(t)|^{2}$ [Eq.(\ref{eq:sup})] and $|\xi _{n}(t)|^{2}$ 
[Eq.(\ref{eq:ap})]. Thus
to fully specify the complex numbers $a_{j}(n)$, it is necessary to
also know their phases, as well as those of the $\xi_{n}(t)$ and
$\psi _{j}(t)$.

To obtain the phases, we note first that it is sufficient to know only the
{\em signs}, since no generality is lost by employing
a {\em real} initial wavefunction $|\Psi (0)\rangle $. The signs can
then be obtained with the help of a simple trick, which we will
illustrate on a generic 2-qubit register state $|\psi\ra = a_0|00\ra +
a_1|01\ra + a_2|10\ra + a_3|11\ra$. Given repeated preparations of this
$|\psi\ra$, we  
perform the following set of measurements:

\begin{itemize}
\item Observation of the two qubits in $|\psi\ra$.
\item A Hadamard transform on the first qubit, followed by observation
of the two qubits. 
\item A Hadamard transform on the second qubit, followed by observation
of the two qubits.
\end{itemize}
The first step yields an estimate of the $|a_i|$. The second one
yields an estimate of $|a_0\pm a_1|$ and $|a_2\pm a_3|$, since under
the Hadamard transform
$|\psi\ra \longmapsto \frac{1}{\sqrt{2}}[(a_0+a_1)|00\ra +
(a_0-a_1)|01\ra + (a_2+a_3)|10\ra + (a_2-a_3)|11\ra ]$. Similarly, the
third step yields an estimate of $|a_0\pm a_2|$ and
$|a_1\pm a_3|$, since $|\psi\ra \longmapsto
\frac{1}{\sqrt{2}}[(a_0+a_2)|00\ra + 
(a_0-a_2)|01\ra + (a_1+a_3)|10\ra + (a_1-a_3)|11\ra ]$. Clearly, this
provides sufficient information for extraction of the signs of all
amplitudes. The generalization to a $\nu$-bit register is obvious:
one performs Hadamard rotations on all $\nu$ qubits. This then yields
$\{|a_0\pm a_1|,|a_2\pm a_3|,|a_4\pm a_5|,...\}$ (after Hadamard on
first qubit);  
$\{|a_0\pm a_2|,|a_1\pm a_3|,|a_4\pm a_6|,...\}$ (after Hadamard on
second qubit); etc. After each Hadamard rotation there are $2^\nu$
coefficients to be estimated. This exponential ``Monte-Carlo scaling'' is the same as
the one we encountered before and is not considered a slow-down for
the reasons detailed above. The additional computational cost is in
the Hadamard rotations, $\nu$ of which must be performed. This does
therefore not affect the efficiency of the algorithm. At the end of
the process, if the whole phase space has been sampled, one is left with $\nu 2^\nu$ absolute values equations,
which contain sufficient information to solve for the signs of all the
amplitudes. In practice one will of course sample only a small
(polynomial) portion of the phase space, and care must then be taken
to obtain sufficient equations of the type above to uniquely determine
the signs of the amplitudes of interest.

\subsection{Repetition}
The steps outlined above generate the energies $\{ E_n \}$ and estimates of 
amplitudes $\{ a_j(n) \}$ needed to perform the sum in
Eq.(\ref{eq:cft}). The whole process must now be repeated many times, on the
order of the required accuracy, in order to complete the
calculation. Due to the speed up in the implementation of
the propagation step, the algorithm performs exponentially faster than
any exact classical algorithm designed to solve the same task.

\section{Discussion and Conclusions}
\label{conclusions}

QCs are still far from being a panacea, and doubts have been
raised whether they will ever replace ordinary, classical
computers.\cite{Unruh:95Haroche:96Warren:97} Such worries are
invariably based on the immense difficulties associated with
maintaining phase coherence throughout the computation, i.e., the
``decoherence problem.'' However, a remarkable theory of quantum error
correction codes has recently 
been constructed,\cite{calderbank:96Steane:96a} in which a ``logical
qubit'' is encoded in the larger Hilbert space of several
physical qubits.\cite{us:DFD} It has been shown that as long as the
error rate is 
sufficiently small, it is possible to perform {\em fault-tolerant}
quantum computation, i.e., the computation can be stabilized and be
made fully robust to
errors.\cite{Shor:96Gottesman:97aPreskill:97aKnill:98} These
advances greatly enhance the prospects of the eventual construction of
useful QCs, beyond the current highly rudimentary prototypes. Building
on these hopes, we have presented here an algorithm for calculating
the thermal rate
constant on a QC. The algorithm involves an initialization step of the
QC into an equal superposition of position eigenstates; a propagation
using an adaptation to QCs of the well-known FFT technique; and
finally, a sequence of measurements yielding the energy spectrum and
amplitudes. Under reasonable assumptions about the distribution of energy
eigenvalues the algorithm runs in polynomial time. The algorithm thus
outperforms any exact classical simulation, which is bound to be
exponential. This clearly demonstrates the potential utility of QCs in
future applications to quantum chemistry problems.

Our approach was somewhat of a ``brute force'' one, in that we did
not attempt to optimize the algorithm using such fruitful concepts as
``direct and correct'' low-rank expressions for the rate
constant.\cite{Miller:98} Such optimizations, while 
ineffectual in altering the essential exponential speed up achieved by
use of a QC, may still be important in practice, especially in the
early stages of the application on a small-scale QC of an algorithm
such as described here. Further work is hence desirable to optimize
the algorithm.

Finally, it would be interesting to check the effect of noise and
other types of errors affecting the evolution of the QC on the present
algorithm. It has been shown, e.g., in the case of the ion trap QC,
that factoring becomes impossible once random phase fluctuations in
the laser pulses exceed a certain threshold.\cite{Miquel:97} We intend
to study similar noise-related issues using numerical simulations in
the context of the present algorithm.

\acknowledgements
We wish to thank Prof. William H. Miller for a critical reading of the
manuscript and many useful comments, and Dr. Christof Zalka for helpful
correspondence. DAL gratefully acknowledges partial support from NSF Grant
CHE 97-32758. HW is supported by the Laboratory Directed
Research and Development (LDRD) project from the National Energy Research
Scientific Computing (NERSC) Center, Lawrence Berkeley National
Laboratory, and also by NSF Grant CHE 97-32758.

\end{multicols}

\end{document}